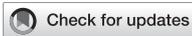

# A highly-compact and ultra-fast homogeneous electromagnetic calorimeter based on oriented lead tungstate crystals

L. Bandiera[1]*, V. G. Baryshevsky[2], N. Canale[1], S. Carsi[3,4], S. Cutini[5], F. Davì[1,6], D. De Salvador[7,8], A. Gianoli[1], V. Guidi[1,9], V. Haurylavets[2], M. Korjik[2], A. S. Lobko[2], L. Malagutti[1], A. Mazzolari[1,9], L. Montalto[10], P. Monti Guarnieri[3,4], M. Moulson[11], R. Negrello[1,9], G. Paternò[1], M. Presti[3,4], D. Rinaldi[11,10], M. Romagnoni[1], A. Selmi[3,4], F. Sgarbossa[8,10], M. Soldani[1,9], A. Sytov[1], V. V. Tikhomirov[2] and E. Vallazza[3]

[1]Istituto Nazionale di Fisica Nucleare, Sezione di Ferrara, Ferrara, Italy, [2]Institute for Nuclear Problems, Belarusian State University, Minsk, Belarus, [3]Istituto Nazionale di Fisica Nucleare, Sezione di Milano Bicocca, Milan, Italy, [4]Dipartimento di Scienza e Alta Tecnologia, Università Degli Studi Dell'Insubria, Como, Italy, [5]Istituto Nazionale di Fisica Nucleare, Sezione di Perugia, Perugia, Italy, [6]Dipartimento di Ingegneria Civile, Edile e Architettura, Università Politecnica Delle Marche, Ancona, Italy, [7]Istituto Nazionale di Fisica Nucleare, Laboratori Nazionali di Legnaro, Legnaro, Italy, [8]Dipartimento di Fisica e Astronomia, Università Degli Studi di Padova, Padua, Italy, [9]Dipartimento di Fisica e Scienze Della Terra, Università Degli Studi di Ferrara, Ferrara, Italy, [10]Dipartimento di Scienze e Ingegneria Della Materia, dell'Ambiente ed Urbanistica, Università Politecnica Delle Marche, Ancona, Italy, [11]Istituto Nazionale di Fisica Nucleare, Laboratori Nazionali di Frascati, Frascati, Italy

Progress in high-energy physics has been closely tied to the development of high-performance electromagnetic calorimeters. Recent experiments have demonstrated the possibility to significantly accelerate the development of electromagnetic showers inside scintillating crystals typically used in homogeneous calorimeters based on scintillating crystals when the incident beam is aligned with a crystallographic axis to within a few mrad. In particular, a reduction of the radiation length has been measured when ultrarelativistic electron and photon beams were incident on a high-$Z$ scintillator crystal along one of its main axes. Here, we propose the possibility to exploit this physical effect for the design of a new type of compact e.m. calorimeter, based on oriented ultra-fast lead tungstate (PWO-UF) crystals, with a significant reduction in the depth needed to contain electromagnetic showers produced by high-energy particles with respect to the state-of-the-art. We report results from tests of the crystallographic quality of PWO-UF samples via high-resolution X-ray diffraction and photoelastic analysis. We then describe a proof-of-concept calorimeter geometry defined with a Geant4 model including the shower development in oriented crystals. Finally, we discuss the experimental techniques needed for the realization of a matrix of scintillator crystals oriented along a specific crystallographic direction. Since the angular acceptance for e.m. shower acceleration depends little on the particle energy, while the decrease of the shower length remains pronounced at very high energy, an oriented crystal calorimeter will open the way for applications at the maximum energies achievable in current and future experiments. Such applications span from forward calorimeters, to compact beam dumps for the search for light dark matter, to source-pointing space-borne $\gamma$-ray telescopes, to decrease the size and the cost of the calorimeter needed to fully contain e.m. showers initiated by GeV to TeV particles.

# 1 Introduction

The key factor in the success of many particle and astroparticle physics experiments has been the use of homogeneous crystal electromagnetic (e.m.) calorimeters to achieve the required resolution for energy measurements. Inorganic scintillator crystals have been widely exploited for the construction of e.m. calorimeters for decades, including for the L3 calorimeter ($Bi_4Ge_3O_{12}$, BGO) at LEP (CERN) [1] and the CMS electromagnetic calorimeter (ECAL) [2] at LHC. Lead tungstate ($PbWO_4$, PWO) scintillator [3], which is used in the CMS ECAL, with over 80,000 crystals [4], has become the most widely used inorganic scintillation material in high-energy physics instrumentation.

In many space-borne experiments, *e.g.*, the Fermi Telescope [5] and DAMPE [6], a crystal e.m. calorimeter is one of the key sub-detectors, designed to perform measurements over a wide range in energy from a few GeV up to several TeV. The higher the material density and the atomic number ($Z$), the shorter the radiation length $X_0$. The radiation length $X_0$, represents the average distance traveled by a high-energy electron before it loses all but $1/e$ of its energy through bremsstrahlung, as well as 7/9 of the average distance for pair production by a high-energy photon (see Figure 1). For this reason, dense, high-$Z$ crystal scintillators (*e.g.*, PWO, BGO, $BaF_2$, *etc.*) were developed to build calorimeters to maintain the detector as compact as possible, since the higher the energy of the primary particle, the larger the propagation length and the volume occupied by the electromagnetic shower. Due to their proven potential, homogeneous crystal calorimeters will be quite useful for future discoveries. Since particle energies for next-generation accelerators will be an order of magnitude higher, calorimeters built with the current technology will have to be longer and more expensive.

At present, the influence of the crystalline lattice of the scintillator on the e.m. shower is usually completely ignored both in calorimeter design and simulation. On the other hand, it is well known that the crystal lattice may strongly modify the physical processes underlying the operation of e.m. calorimeters. In particular, when the trajectory of an $e^{\pm}$ is nearly parallel to either the crystal planes or axes, one can average the Coulomb potential of separate atoms along lattice planes/strings [7]. This planar/axial electric field may reach values of $E = 10^9 – 10^{11}$ V/cm, and at ultra-high energy, the average electric field felt by the particle in its rest frame is Lorentz contracted and can become comparable to the Schwinger critical field, $E_0 = m^2c^3/e\hbar \simeq 1.3 \times 10^{16}$ V/cm, for particles crossing the crystal with a Lorentz factor $\gamma$ of the order of $10^5$–$10^7$, where $m$ is the mass of the particle, $c$ the speed of light, $e$ the electron charge and $\hbar$ the reduced Planck constant. Fields of this strength are characteristic of the atmosphere of a pulsar or a supernova. These crystalline strong fields (SF) stimulate intense hard-photon emission by $e^{\pm}$, as well as intense pair production by photons [7–11]. The resulting enhancement of radiation and pair-production processes accelerates the development of the e.m. shower, which results in a reduction of the $X_0$ and therefore of the shower length (see Figure 2) in comparison with amorphous materials (or randomly oriented crystals).

The angular range for the SF effects can be estimated as

$$\Theta_0 = \frac{U_0}{mc^2}, \quad (1)$$

with $U_0$ the axial potential well depth. For the axes of $PbWO_4$ (PWO), $\Theta_0 \sim 1$ mrad. $\Theta_0$ corresponds to the extent of the angular range over which enhancement of radiation by electrons/positrons and pair production by photons is maximal, but the SF effects are still measurable up to 0.5°–1° of misalignment between the incident particle and the crystal axes. The enhancement of bressmtrahlung and pair production cross sections increases with the initial particle energy. In particular, for PWO it is expected a saturation in the multi-TeV energies; for instance the maximum of the electromagnetic shower initiated by multi-TeV particles (e.g., 10–20 TeV) would be fully contained in about 10 $X_0$, while much more $X_0$s are needed to contain it in case of random orientation of the crystal.

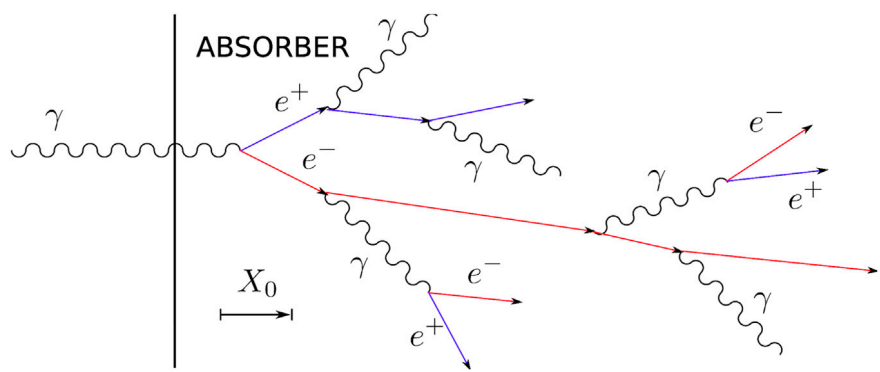

FIGURE 1
Development of an electromagnetic shower initiated by a high-energy photon.

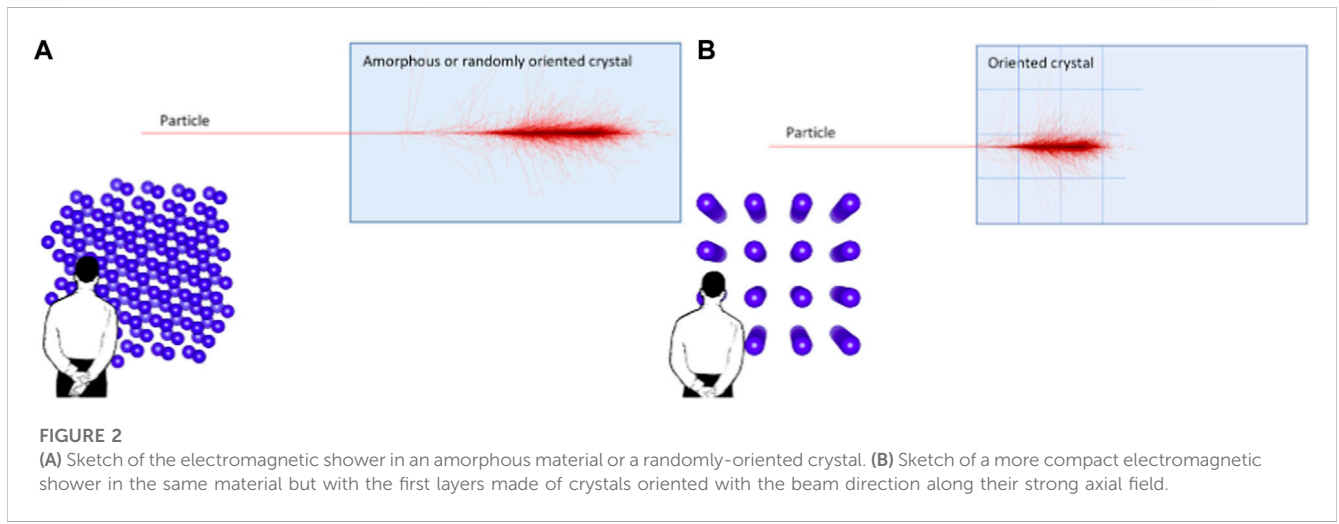

FIGURE 2
(A) Sketch of the electromagnetic shower in an amorphous material or a randomly-oriented crystal. (B) Sketch of a more compact electromagnetic shower in the same material but with the first layers made of crystals oriented with the beam direction along their strong axial field.

This phenomenon has been investigated since the 1980s with single-element crystals such as Si, Ge and W [13,14], and was exploited by the NA48 experiment at CERN to develop a detector with an iridium-crystal photon converter with a reduced $X_0$ [13]. In recent years, experimental tests have shown that a significant reduction in $X_0$ can be achieved with high-Z crystal scintillators such as PWO [15–19].

In this paper, we present a new type of compact e.m. calorimeter, based on oriented ultra-fast lead tungstate (PWO-UF) crystals, with a significant reduction in the depth needed to contain e.m. showers produced by high-energy particles in comparison with currently available devices. In Section 2, we describe the crystal structure of PWO-UF and its characterization via high-resolution X-ray diffraction and photoelastic analysis. In Section 3, we present a new Geant4 model that includes the orientational effects in crystals, which we have used to simulate the shower development in PWO crystals with the aim of choosing the calorimeter geometry. In Section 4 we describe the technology needed to assemble a layer of oriented crystals, while in Section 5 we discuss possible applications in particle and astroparticle physics, such for the construction of compact e.m. calorimeters for fixed target experiments in high-energy physics and very-high-energy (VHE) gamma-ray telescopes in astroparticle physics.

## 2 A new ultra-fast PWO scintillator and its crystallographic characterization

Recently, an improved type of lead tungstate scintillator has been developed [16,20]. Known as PWO-UF (ultra-fast PWO), this scintillator features a scintillation decay time constant of 640 ps and a high radiation tolerance to the electromagnetic component of ionizing radiation, which make it an outstanding candidate for the dual readout of scintillation and Cherenkov photons in electromagnetic calorimetry at future collider experiments. Crystals of PWO-UF were grown at the facilities of Crytur (Czech Republic) with the Czochralski method from platinum crucibles in a neutral atmosphere. Dopant ions of Y and La at a total level of 1500 ppm were added in an initial charge. The melt was corrected to avoid the creation of additional defects by trivalent impurities.

In [15], the crystalline structure and quality of undoped PWO crystals was shown to be suitable for the exploitation of the SF in axial orientation to reduce $X_0$ and shower length. PWO is a high-Z inorganic crystal with $X_0 \approx 8.9$ mm. The PWO lattice has a scheelite-type structure characterized by a tetragonal lattice with constants $a = b = 5.456$ Å and $c = 12.020$ Å. Figure 3A shows a sketch of the PWO structure highlighting the main <100> and <001> axes. Indeed, for the <100> axes of PWO, $U_0 \approx 500$ eV and $\Theta_0 \sim 1$ mrad (Figure 3B). Thus, PWO-UF combines the $X_0$ reduction obtained by aligning conventional PWO with faster light emission, making it a perfect candidate for use in a new type of compact, ultra-fast homogeneous calorimeter for the high-energy and high-intensity frontiers.

Before discussing this possibility, one has to demonstrate that the new PWO-UF has a crystalline quality good enough to allow the SF effect to be exploited. Crystals often have imperfections and their quality can be characterized in terms of the crystal mosaicity. Mosaicity refers to the degree of imperfection or disorder in the alignment of the lattice planes within a crystal. It is due to slight variations in the orientation of the individual crystalline domains, which result in a distribution of lattice orientations of finite width within the overall crystal. If the mosaicity is larger than the angular acceptance of the SF, one may expect a smaller reduction of $X_0$. Another possible drawback would be the presence of residual stresses in the crystalline samples. These could impart a curvature on the lattice planes, thus shifting the orientation of the crystalline axis as a function of position within the sample.

We have performed a surface structural characterization on a sample of PWO-UF using a high-resolution X-ray diffractometer with an 8.04 keV source ($K\alpha_1$ of Cu) in Bragg geometry, in order to obtain information on the surface orientation and superficial mosaicity of the crystal sample [21,22]. Figure 4 shows a series of rocking curves for a 71-mm ($8X_0$) thick PWO-UF sample along the <100> axis over the central 24 mm of the 27 × 27 mm² face of the sample. The rocking curves show the intensity of X-ray diffraction as a function of the angle of incidence of the X-ray beam. The mosaicity of the crystal can be determined by measuring the width of the rocking curve. The FWHM values of the distributions obtained at different points give a mosaicity of ≤ 0.1 mrad, indicating the very good crystallographic quality of the sample (absence of multiple domains). In addition, the axis alignment at different surface positions changes by < 0.2 mrad, demonstrating that is possible to align the entire sample to the direction of beam incidence to within the SF acceptance angle, $\Theta_0$.

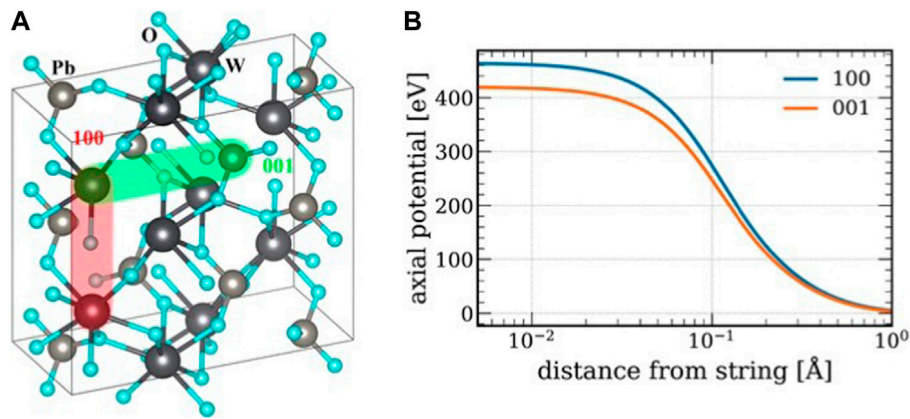

FIGURE 3
Lead tungstate (PWO): Lattice structure (A) with two of the main axes (<100> and <001>) highlighted, whose potentials are also plotted as a function of the distance from the atomic string (B). For a clear visualization of the lattice structure, two neighboring unit cells have been drawn [16].

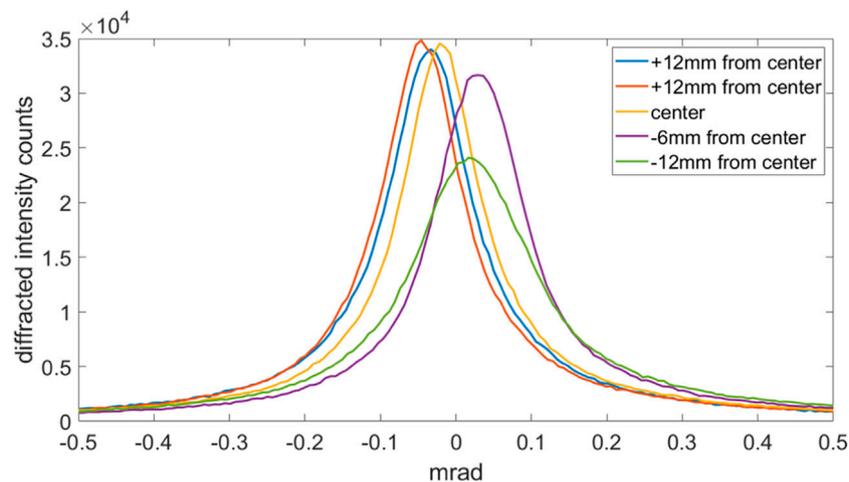

FIGURE 4
Series of rocking curves on the 27 × 27 mm² face of the 71-mm ($8X_0$) thick PWO-UF sample recorded at different positions. High crystal quality is observed, as well as an angular shift of the axis within 0.2 mrad. Measurements performed by high-resolution X-ray diffraction (HR-XRD).

In order to characterize the crystal bulk, the photoelastic technique has been used [23]. For anisotropic crystals, such as PWO, one may use the laser conoscopy technique, in which polarized laser illumination in linear polariscope configurations is used in order to obtain fringe images allowing the quality control of the analyzed sample. The fringe patterns generated by this non-invasive technique carry information about crystallographic orientation and the residual stress/defect conditions, as well as the geometric conditions of the sample. Due to the optical anisotropy of PWO, typical fringe images are produced; stress, misalignments, and geometrical properties can be evaluated locally by analyzing features such as the fringe orders, symmetries and the possible deformation of the pattern shapes. A complete map of the crystal conditions can be obtained by scanning a sample point by point.

A measurement of this type was performed on the $8X_0$ PWO-UF sample with a dedicated optimized apparatus. A map of points at intervals of 2 mm was taken on both face 1 and face 4 of the sample, as illustrated in Figure 5). The experimental setup was prepared in such a way that ≈ 50 $\mu$m of displacement of the fringe positions in the images corresponds to 50 $\mu$rad of misalignment. A fringe pattern has been obtained at each scanned point over the crystal sample; a preliminary analysis shows a misalignment of the axis of less than about 100 $\mu$rad over the whole sample (see Figure 6), in agreement with the XRD data of Figure 4.

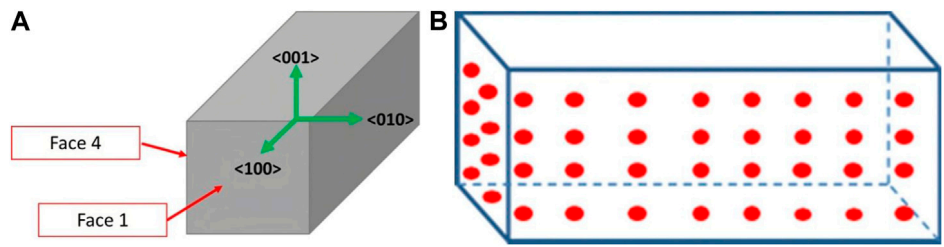

FIGURE 5
**(A)** Geometry of the PWO sample. **(B)** Scheme of the map measurement. A measurement was taken at each point by the laser conoscopy method to evaluate the local crystal conditions.

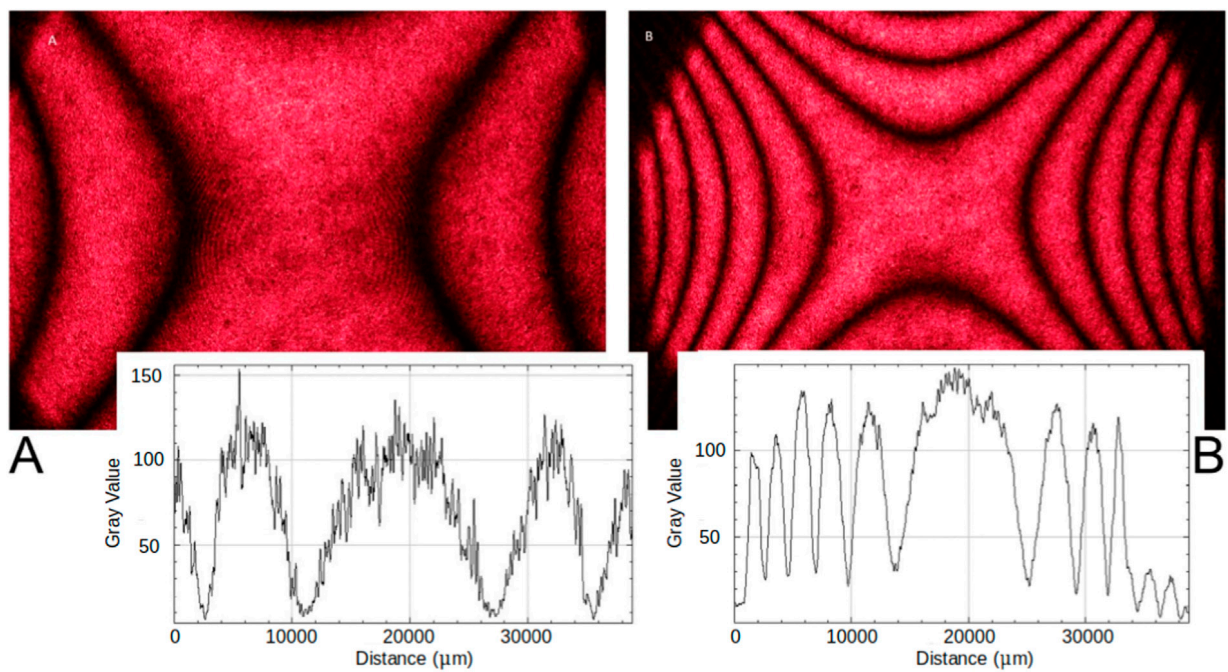

FIGURE 6
Typical fringe patterns obtained for measurement points with the sample observed through face 4 **(A)** and face 1 **(B)** of Figure 5A. The bottom plots for each case show the positions of the interference fringes.

The characterization measurements presented above confirm that the sample of PWO-UF under study is an ideal material for the first prototype of an oriented calorimeter. After these measurements were performed on the test sample as described above, the full suite of measurements was carried out on nine additional crystals from the same manufacturer. The results demonstrate that commercially available PWO-UF crystals grown by the Czochralski method by this manufacturer are of consistently high quality. We emphasize that the suite of described measurements can be carried out for any anisotropic crystalline material as a general quality assessment procedure.

## 3 Design of an oriented calorimeter via Geant4

In order to design an oriented calorimeter, one has to be able to simulate the e.m. shower development inside a crystal under the SF conditions. Since the lattice structure of materials is in general not considered in Geant4 [24], we have developed a new e.m. shower model [25] that includes the enhancement of bremsstrahlung and pair-production in oriented crystals, starting from a previously developed algorithm based on the Baier Katkov quasiclassical method [26–28], which has already been validated vs experiments [7,15,29–31]. In particular, the enhancement in axial configuration was obtained by rescaling the cross sections of bremsstrahlung and pair-production processes according to the results provided by full Monte Carlo computations based on the aforementioned Baier-Katkov method. The correction coefficients for the involved cross-sections were calculated at energies ranging from 1 GeV up to 3 TeV for various materials and axial orientations, assuming a perfect alignment with the considered axis: we chose for simulation <001> axes, being already validated in [15]. During the simulation of an e.m. shower in a crystal, particles undergo bremsstrahlung and pair-production with a probability calculated as if they were perfectly aligned with the selected crystalline axis. Apart from that, they are tracked according to the standard Geant4 mechanism, one by one, down to a kinetic energy of 20 keV and 200 keV, respectively. This approximation works quite well since the dispersion of the secondary particles generated during the shower does not exceed a few mrad. As an example, for 1 TeV photons impinging on a PWO crystal with a thickness of $22X_0$, the fraction of secondary particles generated outside a cone with an aperture of 2 mrad and having an energy higher than 5 GeV is 1%. Therefore, most of the energetic particles, namely, the ones that support the shower, fulfil the condition for which the strong field effects manifest themselves significantly. As a consequence, our model is detailed enough to catch the main contributions to the shortening of the free path length in oriented crystals. We selected as examples of simulations the range of interest for the applications mentioned above (e.g., HIKE ([32] or a γ-ray satellite), i.e., 20, 120, 1000 and 3,000 GeV. The first two energies are also available for bench tests at the extracted beam lines of the CERN Super Proton Synchrotron [15,33]. In particular, 20 GeV corresponds approximately to the energy at which the Lorentz-contracted electric field of the crystal axis reaches the Schwinger value $E_0$.

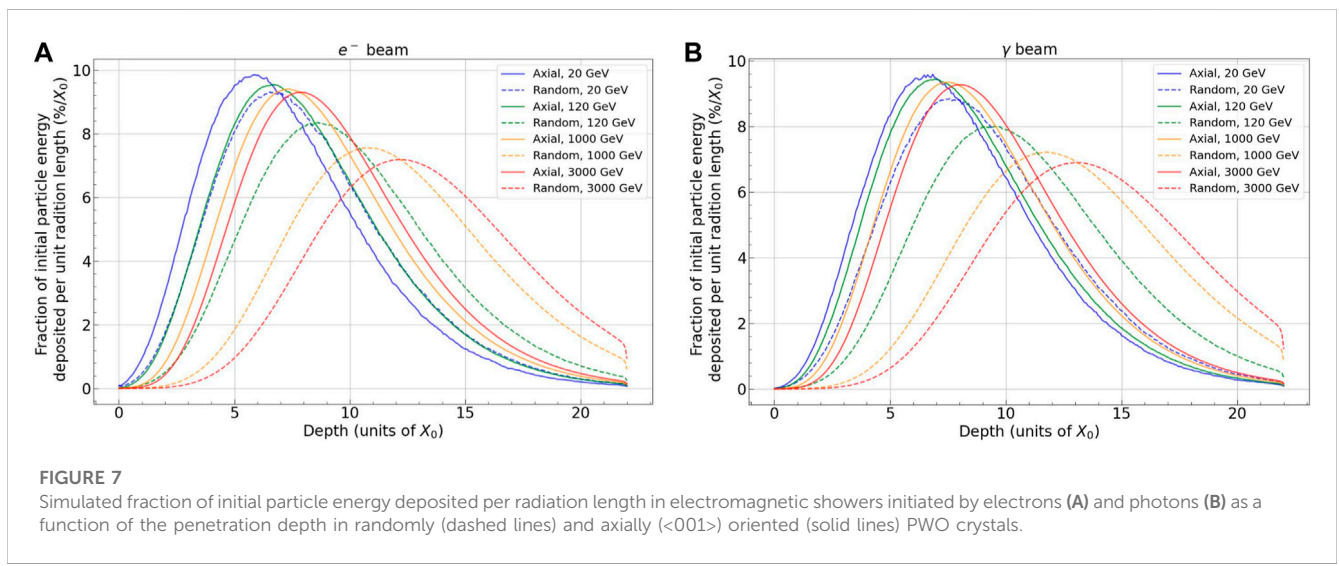

FIGURE 7
Simulated fraction of initial particle energy deposited per radiation length in electromagnetic showers initiated by electrons **(A)** and photons **(B)** as a function of the penetration depth in randomly (dashed lines) and axially (<001>) oriented (solid lines) PWO crystals.

Figure 7 presents the simulated fraction of energy deposited per radiation length in electromagnetic showers initiated by electrons (A) and photons (B) as a function of the penetration depth in (dashed-lines) randomly and (solid line) axially (<001>) oriented PWO: when the crystal is axially oriented, the shower development is accelerated, with the consequence that the energy deposit peaks are generally closer to the front surface of the sample. Moreover, when on axis, the curves from 20 GeV to 3,000 GeV have maxima much closer to each other than in the random case. For example, 20 GeV (3,000 GeV) electrons feature a shower peak at ~ $6.5X_0$ (~ $12.1X_0$) when randomly oriented and at ~ $5.9X_0$ (~ $7.9X_0$) when axially oriented [17]. The integrals of the curves in Figure 7, i.e., the total fractions of the primary energy deposited in the samples up to a certain shower depth, are shown in Figure 8. It is evident that the difference between the axial and the random curves increases with initial particle energy, which highlights how the strength of the strong-field enhancement grows with the energy. From Figure 8, we may notice that for either electrons or photons in the full energy range, for depth $\lesssim 2X_0$, the energy deposit consists of $\leq 1\%$ of the incident particle energy: the absolute difference between the total energy deposited in the on-axis and random configurations is small, whereas the ratio between them is the largest. Between ~ $2X_0$ and the position of the energy deposition peak in random configuration (e.g., $6.5X_0$ and $7.6X_0$ at 20 GeV and $12.1X_0$ and $13.2X_0$ at 3,000 GeV, for electrons and photons, respectively), the energy deposited per $X_0$ corresponds to a significant part of the primary energy, and the difference between the total energy deposited in the axial and random configurations is proportional to the penetration depth as presented in Figure 9. Figure 9 shows the difference between the percentage of deposited energy in PWO crystals in axial and random orientation as a function of the penetration depth: results at different energies for either electrons (A-left) or photons (B-right) are presented. At depths beyond the position of the peak in the energy deposition curve (Figure 7) for the random case, the axial-random difference in energy deposition decreases. Finally, at a thickness of $\gtrsim 20X_0$, almost all the initial energy has been deposited in the crystal regardless of the lattice orientation.

Summarising, Figures 7–9 would prove useful in selecting the optimal crystal thickness to study the shower development at different stages for different energy ranges and/or applications.

## 4 How to build a layer of oriented crystals

One of the biggest challenges in designing a compact oriented calorimeter is to develop a mechanical structure to arrange the crystals in a matrix and keep them aligned to within less than half of the maximum of the SF angle, $\Theta_0$ (≈0.5 mrad for PWO), while accounting for temperature fluctuations and handling stresses. The crystals may also have miscut angles between crystallographic and morphological surfaces, which must be accounted for. Starting with commercially available PWO-UF ingots, the first step is to measure the miscut angle, *i.e.,* the angular difference between the actual surface and the desired lattice orientation, with high-resolution X-ray diffraction (HR-XRD). The ingots may then be precisely cut in the desired orientation. As seen in Section 3, the acceleration of the e.m. shower is maximal in the first few radiation lengths, where the particles have the highest energies and the smaller angular spread. For a longitudinally segmented crystal calorimeter, depending on the application and cell dimensions, it may be possible to use oriented crystals only for the first layer or layers, i.e., up to the depth at which the probability for an incident e.m. particle to reach without interactions is negligible for the application at hand. In any event, a good cost-effective compromise for a proof-of-concept instrument could be a two-layer, longitudinally segmented calorimeter with a first layer depth of $\approx 5X_0$ (from preliminary simulation) and second, unaligned layer $\approx 15X_0$ long. About 98% of incident high-energy photons would be expected to convert in the first layer, while as seen from Figure 7 to Figure 8, the fraction of energy deposited in the first layer is expected to be significantly increased in the axial orientation at typical test-beam energies of 20–120 GeV.

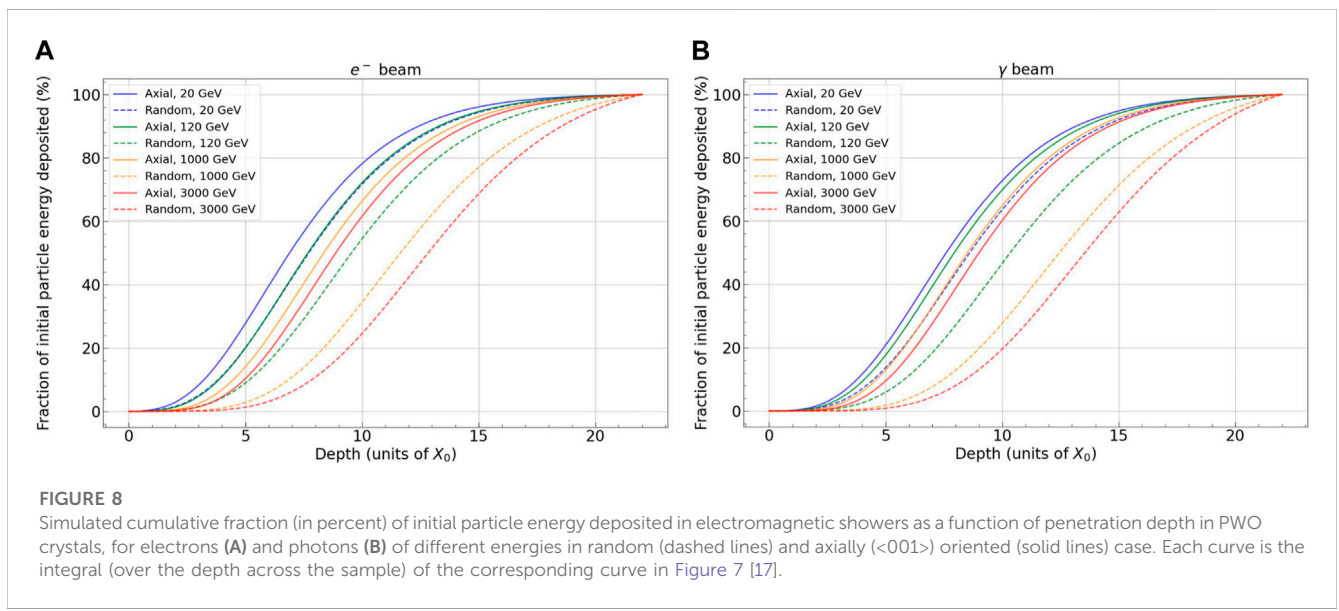

FIGURE 8
Simulated cumulative fraction (in percent) of initial particle energy deposited in electromagnetic showers as a function of penetration depth in PWO crystals, for electrons (A) and photons (B) of different energies in random (dashed lines) and axially (<001>) oriented (solid lines) case. Each curve is the integral (over the depth across the sample) of the corresponding curve in Figure 7 [17].

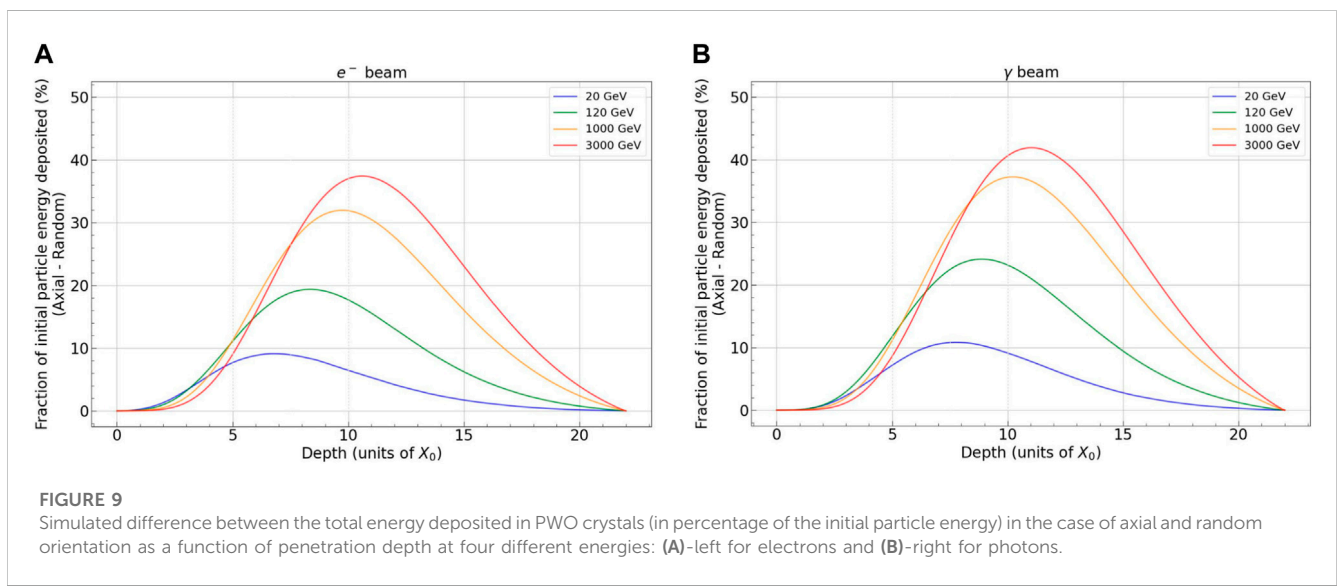

FIGURE 9
Simulated difference between the total energy deposited in PWO crystals (in percentage of the initial particle energy) in the case of axial and random orientation as a function of penetration depth at four different energies: (A)-left for electrons and (B)-right for photons.

The calorimeter design may consist of a matrix of crystals coupled to photo-sensors [16], as shown in Figure 10. This illustration is only an example of a design that can be modified and adapted to different applications. For applications requiring maximum detection efficiency and minimization of rare fluctuations (for example, rare-particle searches), it may be desirable to align all layers.

Special attention must be devoted to the realization of the oriented crystal layers. For the prototype illustrated in Figure 10, the relative orientation between the 9 crystals will be obtained by an improved version of the specially developed procedure employed in [34]. To ensure a precise angular positioning of the atomic planes and axes of the crystals, the miscut angle will be characterized using a custom-made setup based on laser autocollimators and high-resolution X-ray diffraction. Once the miscut angle is characterized and the crystal lateral surfaces coated using a reflective paint, the sample front faces will be aligned through a set of linear and angular adjusters and glued together (Figure 10A). Glue will also be used as a spacer material between the joined faces of the crystals, as its thickness along the faces of the crystals can be manipulated in order to generate a wedge to align the crystalline structures using the physical surfaces as references. The quality of the alignment will be verified through Fizeau interferometry and high-resolution X-ray diffraction. The method described can be adapted to large calorimeters with hundreds of crystals, such as that for the HIKE experiment (see Section 5).

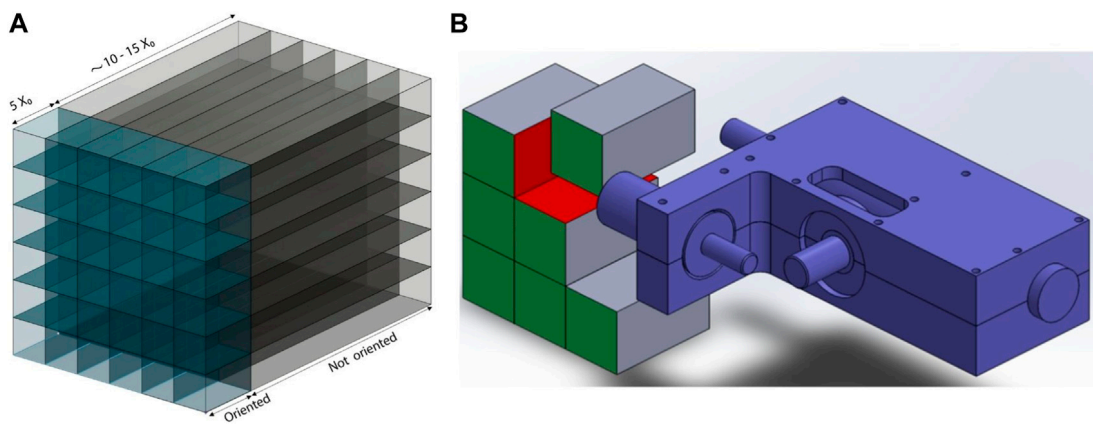

FIGURE 10
Conceptual sketch of an oriented crystal calorimeter matrix **(A)**. Sketch of the assembly tool **(B)**.

A standard technique to increase the efficiency and hermeticity of a crystal calorimeter is to align the crystal volumes such that the gaps between adjacent crystals do not point in the expected direction of particle incidence. By cutting the crystals such that the desired axis points in the direction of particle incidence while the geometrical surfaces do not, an oriented-crystal calorimeter can easily be designed with a non-projective geometry.

## 5 Discussion and applications

An oriented crystal calorimeter would find various applications in particle and astroparticle physics. In particle physics, it can reduce the calorimeter volume and enhance the performance in fixed-target experiments, which are intrinsically forward (within a few mrad). In particular, the calorimeter depth can be reduced proportionally to the reduced e.m. shower length, making the calorimeter relatively insensitive to hadronic interactions, allowing it to detect photons in a neutral hadron beam without being blinded by the beam itself. This method will find immediate application in the HIKE-KLEVER [32] experiment at the CERN SPS. In the third phase of the proposed HIKE program, the KLEVER experiment will measure the branching ratio for the ultra-rare decay $K_L \to \pi^0 \nu \bar{\nu}$, which is predicted in the Standard Model to be about $3 \times 10^{-11}$. The primary background is from $K_L \to \pi^0 \pi^0$ decays (BR = $8.64 \times 10^{-4}$) with two lost photons. Hermetic, high-performance photon vetoes are essential for the reduction of this background by eight or more orders of magnitude. The HIKE small-angle calorimeter (SAC) is a very-forward veto calorimeter that sits squarely in neutral beam to intercept photons that might otherwise escape through the beam exit. It must detect and veto the photons from $K_L \to \pi^0 \pi^0$ decays while maintaining insensitivity to more than 500 MHz of neutral hadrons in the beam. This is an ideal case for leveraging the coherent interactions of high-energy photons in oriented crystals to induce prompt electromagnetic showering. The SAC acceptance for photons from $K_L \to \pi^0 \pi^0$ decays in the fiducial volume extends out to no more than ±2 mrad [32]; these photons are in principle inside the SF acceptance. Aligning the SAC crystals will then provide the advantage of the reduced ratio between $X_0$ and the nuclear interaction length, $\lambda_{int}$, thus maintaining the sensitivity of the detector to photons and electrons at reduced physical thickness and thus reduced sensitivity to hadrons in the beam, reducing overall rates in the detector to avoid blinding by hadronic interactions. The fast decay time of PWO-UF will additionally lead to excellent time resolution and double-pulse discrimination, which will further help to reduce the accidental veto rate. The excellent radiation resistance of PWO-UF makes it a good candidate for use in an in-beam veto.

Another application could be related to beam-dump experiments for light dark matter searches, such as NA64 [35], to reduce the dump length. If a dark photon is created in a shower initiated by an electron, it can be detected only if it survives for the remaining dump length. The shorter the length, the higher the sensitivity. The use of oriented crystals for these fixed-target applications is ideal, because the high-energy particles to be detected impinge at well-defined angles and there is virtually no material upstream of the crystals to initiate showers or scatter incident particles. Other applications, for example, in detectors for collider experiments at very forward angles for which particles arrive directly from the interaction point with very little angular spread, may also be considered. Clearly, the potential improvement from the use of aligned crystals in the energy resolution at a given (presumably limited) calorimeter depth depends on the details of the application at hand, notably, on the strength of the constraints on the incident particle trajectories. However, it may be useful to take alignment effects into consideration in the design of any calorimeter. Depending on the production method, crystals may naturally be produced with a crystal axis in rough alignment with the geometric axis [36]; the degree of this alignment may vary from crystal to crystal. If crystal alignment is not to be exploited for a given application, it may be desirable to align the crystals so that no axis is near to the direction of expected particle incidence. If a particle finds itself accidently aligned with a crystal axis, the energy deposition profile will be different from that of other crystals, leading to potential contributions to the energy resolution from intercalibration effects. These effects are likely more important for calorimeters of limited depth, or with longitudinal segmentation.

Applications of oriented crystals in VHE astrophysics would be a pioneering approach that would open new possibilities for gamma-ray detection technology on satellites. In case of pointing strategy (a possibility already attainable by the Fermi-LAT telescope), an oriented-crystal based hodoscopic calorimeter would enhance the sensitivity of the telescope above a few GeV, thanks to the increase of pair production cross section in the source direction. In addition, the electron/positron electromagnetic showers initiated by gamma-rays with energies even larger than 100 GeV would be contained inside a smaller volume than for standard detectors. For instance, in Figure 7 it is shown that at 3 TeV the maximum of the shower is fully contained in about 8 $X_0$, which is less than the full length of the Fermi-LAT (10.1 $X_0$ for the tracker + calorimeter). Since for space missions it is mandatory to minimize the dimension and the weight of the detectors on-board, a gamma-ray satellite based on a compact oriented calorimeter could have some advantages. For instance, with the same weight as a standard satellite, the reduction of $X_0$ would make it possible to reduce the longitudinal dimension and, at the same time, increase the transversal area of the detector, thus increasing its sensitivity. Indeed, in VHE astrophysics, the primary challenge is the very limited photon flux. A larger area combined with a better containment of the shower in the 100–300 GeV range, could be quite useful since this energy range is not well covered either from space or from ground detectors such as future IACTs (e.g., CTA) [11]. Furthermore, the higher sensitivity in one specific direction would naturally reduce the influence of the isotropic background to better identify the gamma-ray source. Since the maximum of the shower enhancement in crystals occurs within a few mrad (1–2 mrad for PWO [15]), one may exploit this anisotropy to improve the angular resolution and therefore the localization of the sources, without using a complex converter-tracker system. Such an apparatus would continue to operate in a standard way when operated without pointing. Several fields of astrophysics could be explored using the pointing strategy, for example, observation of unidentified FERMI gamma-ray sources, follow-up of transient and multi-messenger sources and pointing of the galactic center for dark matter discrimination (for either the Milky Way or other specific galaxies as dwarf spheroidal galaxies).

## 6 Conclusion

Here we propose a new type of compact e.m. calorimeter based on oriented, ultra-fast PWO (PWO-UF) scintillator crystals, with a key feature of a reduced depth needed to contain the e.m. shower produced by high-energy particles with respect to the state-of-the-art. This reduction is caused by an acceleration of the e.m. shower development when a photon/electron beam with energy of several GeV or more is aligned with one of the main PWO lattice axes.

Since the reduction of the radiation length is strongest when the angle between the beam direction and the crystallographic axis is within $\Theta_0 \approx 1$ mrad, we have verified that the PWO-UF crystals grown by the Czochralski method available from Crytur are of high enough quality for use in an oriented-crystal calorimeter, specifically, with regards to the mosaic spread, which ideally should be much smaller than $\Theta_0$. For our samples, we measured via HR-XRD a superficial mosaicity of ≤0.1 mrad and a negligible change in the axis alignment at different surface positions. The latter was confirmed by photoelastic analysis for the entire crystal depth. A new Geant4 model including the shower development in oriented crystals has been used to define a possible calorimeter geometry for future tests and applications. As a proof-of-concept, one may realize a longitudinally segmented calorimeter with only the first layer of 5–7.5 $X_0$ oriented, resulting in a significant reduction in shower length. Finally, we have introduced a technique based on optical alignment to realize a matrix of axially aligned PWO-UF crystals that can be employed in realistic, large calorimeters.

In conclusion, we have presented a-proof-of-concept design of an ultra-fast, compact oriented calorimeter which can be the basis for various applications, including forward calorimeters, compact beam dumps for the search for light dark matter, and source-pointing space-borne γ-ray telescopes, which can reduce the size and cost of calorimeters needed to contain the e.m. showers initiated by particles ranging from GeV to TeV energies.

## Data availability statement

The raw data supporting the conclusion of this article will be made available by the authors, without undue reservation.

## Author contributions

LB: Conceptualization, Funding acquisition, Investigation, Methodology, Project administration, Resources, Software, Supervision, Writing–original draft, Writing–review and editing. VB: Resources, Writing–review and editing. NC: Data curation, Visualization, Writing–original draft, Writing–review and editing. SC: Investigation, Methodology, Writing–review and editing. FD: Investigation, Methodology, Resources, Writing–review and editing. DD: Conceptualization, Investigation, Methodology, Project administration, Resources, Writing–review and editing. AG: Investigation, Methodology, Resources, Writing–review and editing. VG: Funding acquisition, Resources, Supervision, Writing–review and editing. VH: Conceptualization, Investigation, Methodology, Software, Writing–review and editing. MK: Methodology, Writing–original draft, Writing–review and editing. AL: Funding acquisition,

Investigation, Methodology, Writing–review and editing. LMa: Methodology, Writing–review and editing. AM: Conceptualization, Funding acquisition, Investigation, Methodology, Resources, Supervision, Writing–original draft, Writing–review and editing. LMo: Conceptualization, Data curation, Formal Analysis, Investigation, Methodology, Resources, Validation, Visualization, Writing–original draft, Writing–review and editing. PM-G: Conceptualization, Investigation, Methodology, Software, Writing–review and editing. MM: Conceptualization, Funding acquisition, Investigation, Methodology, Project administration, Resources, Software, Supervision, Visualization, Writing–original draft, Writing–review and editing. RN: Data curation, Visualization, Writing–original draft, Writing–review and editing. GP: Investigation, Software, Writing–review and editing. MP: Conceptualization, Funding acquisition, Investigation, Methodology, Project administration, Resources, Supervision, Writing–original draft, Writing–review and editing. DR: Funding acquisition, Investigation, Methodology, Resources, Supervision, Writing–original draft, Writing–review and editing. MR: Conceptualization, Data curation, Formal Analysis, Investigation, Methodology, Validation, Writing–original draft, Writing–review and editing. ASe: Data curation, Investigation, Methodology, Writing–review and editing. FS: Investigation, Methodology, Writing–review and editing. MS: Conceptualization, Data curation, Formal Analysis, Investigation, Methodology, Software, Validation, Writing–original draft, Writing–review and editing. ASy: Investigation, Methodology, Software, Writing–review and editing. MT: Visualization, Writing–review and editing. VT: Conceptualization, Investigation, Methodology, Project administration, Resources, Software, Supervision, Writing–original draft, Writing–review and editing. EV: Conceptualization, Investigation, Methodology, Project administration, Resources, Supervision, Writing–review and editing.

# Funding

The author(s) declare financial support was received for the research, authorship, and/or publication of this article. This work was supported by INFN CSN5 (OREO and MC-INFN projects) and CSN1 (NA62, RD_FLAVOUR and RD_MUCOL projects) and the European Commission through the H2020-INFRAINNOV AIDAINNOVA (G. A. 101004761), H2020-MSCA-RISE N-LIGHT (G. A. 872196) and EIC-PATHFINDER-OPEN TECHNO-CLS (G. A. 101046458) projects.

# Acknowledgments

ASy acknowledges support from the H2020-MSCA-IF-Global TRILLION (G. A. 101032975).

# Conflict of interest

The authors declare that the research was conducted in the absence of any commercial or financial relationships that could be construed as a potential conflict of interest.

# Publisher's note

All claims expressed in this article are solely those of the authors and do not necessarily represent those of their affiliated organizations, or those of the publisher, the editors and the reviewers. Any product that may be evaluated in this article, or claim that may be made by its manufacturer, is not guaranteed or endorsed by the publisher.